# Manipulating magnetism and transport properties of EuCd$_2$P$_2$ with a low carrier concentration

Xiyu Chen,[1] Ziwen Wang,[1] Zhiyu Zhou,[1] Wuzhang Yang,[2,3] Yi Liu,[4] Jia-Yi Lu,[5] Zhi Ren,[2,3] Guang-Han Cao,[5,6] Fazel Tafti,[7] Shuai Dong,[1,†] and Zhi-Cheng Wang[1,‡]

[1]*Key Laboratory of Quantum Materials and Devices of Ministry of Education, School of Physics, Southeast University, Nanjing 211189, China*

[2]*School of Science, Westlake University, Hangzhou 310024, China*

[3]*Institute of Natural Sciences, Westlake Institute for Advanced Study, Hangzhou 310024, China*

[4]*Department of Applied Physics, Zhejiang University of Technology, Hangzhou 310023, China*

[5]*School of Physics, Interdisciplinary Center for Quantum Information and State Key Laboratory of Silicon and Advanced Semiconductor Materials, Zhejiang University, Hangzhou 310058, China*

[6]*Collaborative Innovation Centre of Advanced Microstructures, Nanjing University, Nanjing 210093, China*

[7]*Department of Physics, Boston College, Chestnut Hill, Massachusetts 02467, USA*

**ABSTACT**

Materials that exhibit strongly coupled magnetic order and electronic properties are crucial for both fundamental research and technological applications. However, finding a material that not only shows remarkable magnetoresistive responses but also has an easily tunable ground state remains a challenge. Here, we report successful manipulation of the magnetic and transport properties of EuCd$_2$P$_2$, which is transformed from an A-type antiferromagnet ($T_N$ = 11 K) exhibiting colossal magnetoresistance into a ferromagnet ($T_C$ = 47 K) with metallic behavior. The dramatic alteration results from a low hole concentration of $10^{19}$ cm$^{-3}$ induced by changing the growth conditions. Electronic structure and total energy calculations confirm the tunability of magnetism with a small carrier concentration for EuCd$_2$P$_2$. It is feasible to switch between the magnetic states by using field-effect to control the carrier density, thereby changing the magneto-electronic response. The controllable magnetism and electrical transport of EuCd$_2$P$_2$ make it a potential candidate for spintronics.

---

† sdong@seu.edu.cn
‡ wzc@seu.edu.cn

**I. INTRODUCTION**

In recent decades, the utilization of the charge or spin of electrons has led to significant developments in electronic devices for information storage, processing, and communications [1–3]. However, the independent control of charge or spin cannot meet the increasing requirements of ultra-fast response, high integration, and high energy efficiency in newly emerging applications [4–7]. Therefore, it is essential to explore materials that exhibit strong coupling between charge transport and magnetic order [8]. Various efforts have been made to connect these two distinct properties, such as the study of intrinsic ferromagnetic semiconductors, diluted magnetic semiconductors, and magnetic two-dimensional (2D) materials [9–12]. But so far, materials that simultaneously feature tunable magnetic order and electrical transport are scarce.

In a recent report, we demonstrated an enormous colossal magnetoresistance (CMR) in a layered Eu-based phosphide, $EuCd_2P_2$, with A-type antiferromagnetic (AFM) order at 11 K. And the magnitude of CMR rivals that in optimized thin films of manganates [13]. Therefore, $EuCd_2P_2$ holds promise for AFM spintronics at low temperatures due to its strong magnetoresistive response. In this study, we found that the ground state of $EuCd_2P_2$ can be tuned into ferromagnetic (FM) simply by changing the growth conditions. FM-$EuCd_2P_2$ ($T_C$ = 47 K) exhibits metallic behavior, and the resistivity is two orders of magnitude lower than that of AFM-$EuCd_2P_2$ in the region of strong CMR (10 ~ 30 K). The dramatic changes in the magnetic state and transport properties are attributed to a low hole concentration on the order of $10^{19}$ cm$^{-3}$ induced by vacancies in the material. Based on the tunability of $EuCd_2P_2$, we propose controlling its magnetism and charge transport by varying the carrier concentration using an electrostatic gate, which has been applied to manipulate the magnetism of 2D materials [14].

$EuCd_2P_2$ is an exceptional material among Eu-based materials for its magneto-electronic coupling, layered structure, and suitability for the fabrication of thin films or heterojunctions. Previously, the magnetic order of $EuCd_2As_2$, which is isostructural to $EuCd_2P_2$, was tuned from an AFM state to a FM state with a lower $T_C$ of 26 K [15,16]. However, the ground state alteration of $EuCd_2As_2$ did not lead to an explicit change in charge transport. Compared to the representative magnetic semiconductors Eu$X$ ($X$ = O, S, Se, Te), $EuCd_2P_2$ single crystals are much easier to grow using the flux method below 1000°C [9]. Thin films of $EuCd_2P_2$ and heterostructures are also likely to be fabricated due to its layered structure. Moreover, it's promising to further improve the $T_C$ of $EuCd_2P_2$ with carefully controlled carrier density through chemical doping or electrostatic gating.

**II. METHODS**

*Crystal growth.* Single crystals of FM-EuCd$_2$P$_2$ were grown using a molten salt flux method. A total mass of 1 g of Eu ingots (99.999%), Cd ingots (99.999%), and red P lumps (99.999%) were weighed in the mole ratio of 1: 2: 2, then mixed with 4 g of an equimolar mixture of NaCl (99.99%) and KCl (99.99%) in a glove box. The reactants and salts were loaded into a quartz tube directly and sealed under high vacuum (<10$^{-2}$ Pa). Next, the first sealed quartz tube was put into another quartz tube with slightly larger diameter, which was sealed under vacuum as well. The nested tubes were heated to 469°C over 24 h, held for 24 h, then heated to 597°C in a rate of 20°C/h and held for another 24 h. Afterwards, the tubes were heated to 847°C and held for 80 h, and subsequently cooled to 630°C at 1.5°C/h. Millimeter-sized shiny thin flakes of EuCd$_2$P$_2$ single crystals were recovered after washing the product with deionized water. It should be noted that a small diminution of the proportion of Eu (such as Eu: Cd: P = 0.75: 2: 2) has little impact on the composition of final FM-EuCd$_2$P$_2$, manifested by the similar magnetic and transport properties of the single crystals. The crystal growth procedure for AFM-EuCd$_2$P$_2$ was described elsewhere [13].

*Structure Determination and Chemical Compositions Characterization.* To determine the structure of FM-EuCd$_2$P$_2$, single crystals were examined at 150 K via a Bruker D8 Venture diffractometer equipped with I$\mu$S 3.0 Dual Wavelength system (Mo $K\alpha$ radiation, $\lambda = 0.71073$ Å), and an APEX-II CCD detector. The collected frames were reduced and corrected using the Bruker SAINT software. The initial structural model was developed with the intrinsic phasing feature of SHELXT and a least-square refinement was performed using SHELXL2014 [17]. We also examined the unit cell along crystallographic $c$ axis at room temperature on a PANalytical x-ray diffractometer with the Cu $K\alpha$1 radiation ($\lambda = 1.54178$ Å).

*Magnetization and Transport Measurements.* The direct-current (dc) magnetization was measured using a Quantum Design Magnetic Property Measurement System (MPMS-3) with a crystal whose dimensions were about $1.0 \times 0.5 \times 0.03$ mm$^3$ (~0.1 mg). The anisotropic electrical resistivity measurements were carried out on a Quantum Design Physical Property Measurement System (PPMS-9) adopting a standard four-probe technique.

*First-principles calculations.* DFT calculations were performed using Vienna *ab initio* Simulation Pack (VASP) [18]. The electron-ion interactions are described by projector augmented wave (PAW) pseudopotentials [19]. The plane-wave cutoff energy is fixed as 318 eV. The exchange-correlation functional is treated using Perdew-Burke-Ernzerhof-revised (PBEsol) parametrization of the generalized gradient approximation (GGA) [20]. For FM-EuCd$_2$P$_2$, a Γ-centered 11 × 11 × 7 Monkhorst-Pack *k*-mesh is adopted for Brillouin zone sampling. For AFM-EuCd$_2$P$_2$, the hexagonal unit cell is doubled along the *c* axis with a 11 × 11 × 3 *k*-mesh. To better describe the correlation effect, an effective Hubbard correlation *U* value of 5.0 eV using the Dudarev approach is applied to Eu 4*f* orbitals, as used in previous EuCd$_2$As$_2$ system [21]. Both the lattice constants and atomic positions are fully optimized iteratively until the total energy and the Hellmann-Feynman force on each atom are converged to 10$^{-6}$ eV and 0.01 eV/Å, respectively. The Eu vacancy effects was simulated in a 3 × 3 × 2 supercell (90 atoms).

## III. RESULTS AND DISCUSSION

### A. Structural analysis

The crystal structure of EuCd$_2$P$_2$ is illustrated in Fig. 1(a), composed of alternating Eu atom layers and Cd$_2$P$_2$ sheets of edge-sharing tetrahedra, which is known as the CaAl$_2$Si$_2$-type structure (space group $P\bar{3}m1$). Triangular Eu$^{2+}$ (4*f*$^7$) lattice is responsible for the magnetism of EuCd$_2$P$_2$ [13]. Single crystals of FM-EuCd$_2$P$_2$ were grown using a molten salt mixture (NaCl/KCl), similar to the growth of FM-EuCd$_2$As$_2$ [15]. The typical morphology of FM-EuCd$_2$P$_2$ is shown in Fig. 1(b), with dimensions of 1.0 × 0.5 × 0.03 mm$^3$. Unlike the bulky AFM-EuCd$_2$P$_2$ crystals grown via Sn flux method, the thin flakes of FM-EuCd$_2$P$_2$ crystals are brittle.

Room temperature x-ray diffraction (XRD) was used to examine the phase purity of the single crystals, as shown in Fig. 1(c). The intensity of the diffraction pattern is scaled logarithmically to enhance the visibility of the weak peaks, and only (00*l*) reflections of EuCd$_2$P$_2$ are observed. The *c* parameter is calculated to be 7.176(1) Å based on the measured (00*l*) diffractions, which is close to the *c* parameter of AFM-EuCd$_2$P$_2$. The diffraction peaks are highly sharp, with a full width at half maximum (FWHM) as small as 0.04°, indicating the high quality of the single crystals. Single crystal XRD data was also collected at 150 K, and the refinement analysis shows that the structural parameters of FM-EuCd$_2$P$_2$ are much the same as those of AFM-EuCd$_2$P$_2$, except for a tiny amount of Eu vacancies (~0.2%) in the lattice, reminiscent of the case of FM-EuCd$_2$As$_2$. Additional structural details of FM-EuCd$_2$P$_2$ are provided in Tables S1 and S2 in the Supplemental Material (SM) [22].

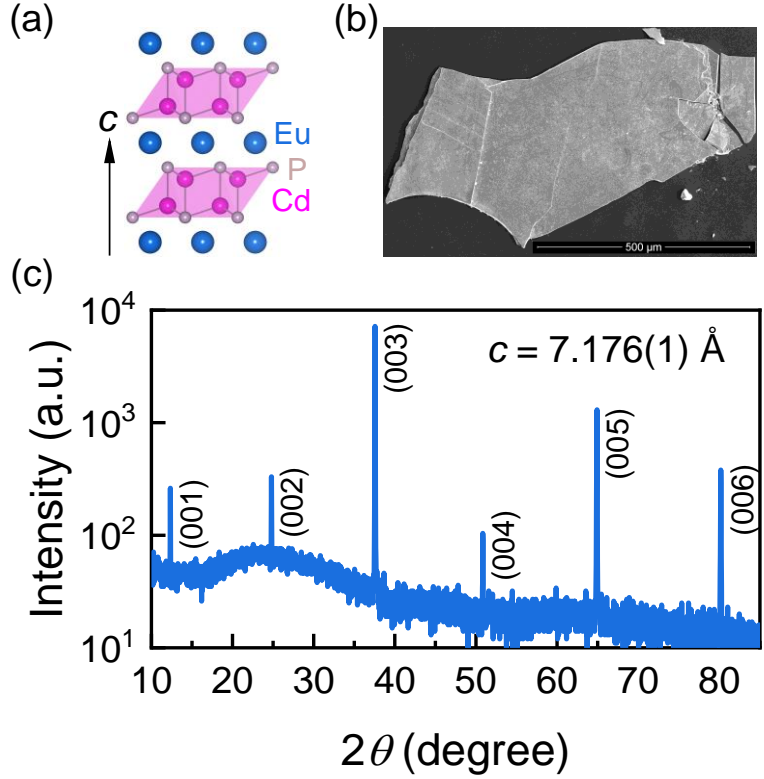

FIG. 1. (a) Crystal structure of $EuCd_2P_2$ projected along the [100] direction. (b) A typical image of FM-$EuCd_2P_2$ crystal grown with salt flux. (c) XRD pattern ($\theta$-$2\theta$ scan) of the FM-$EuCd_2P_2$ single crystal, showing only (00$l$) reflections when the sample was placed on the holder horizontally.

**B. Magnetic properties**

Figure 2(a) shows the isothermal magnetization loops of $EuCd_2P_2$ grown via salt flux at 1.8 K. The in-plane $M(H)$ curve saturates at approximately 0.07 T, which is significantly smaller than the saturation field of 1.6 T observed for the out-of-plane $M(H)$ curve. The ratio of saturation fields achieves as high as 20 times. The strong magnetocrystalline anisotropy suggests that the Eu layer (the $ab$ plane) is the magnetic easy plane. Similarly, AFM-$EuCd_2P_2$ grown by Sn flux also exhibits a large magnetocrystalline anisotropy (~10 times), and the saturation field along the $c$ axis is 1.6 T as well [13], indicating that the interlayer coupling is weak and has little effect on the spin reorientation in the field. The right inset in Fig. 2(a) shows the $M(H)$ curves under low fields, revealing a small but clear hysteresis loop at 1.8 K with a coercivity of 0.55 mT. These observations suggest that $EuCd_2P_2$ (salt flux) is an extremely soft FM material.

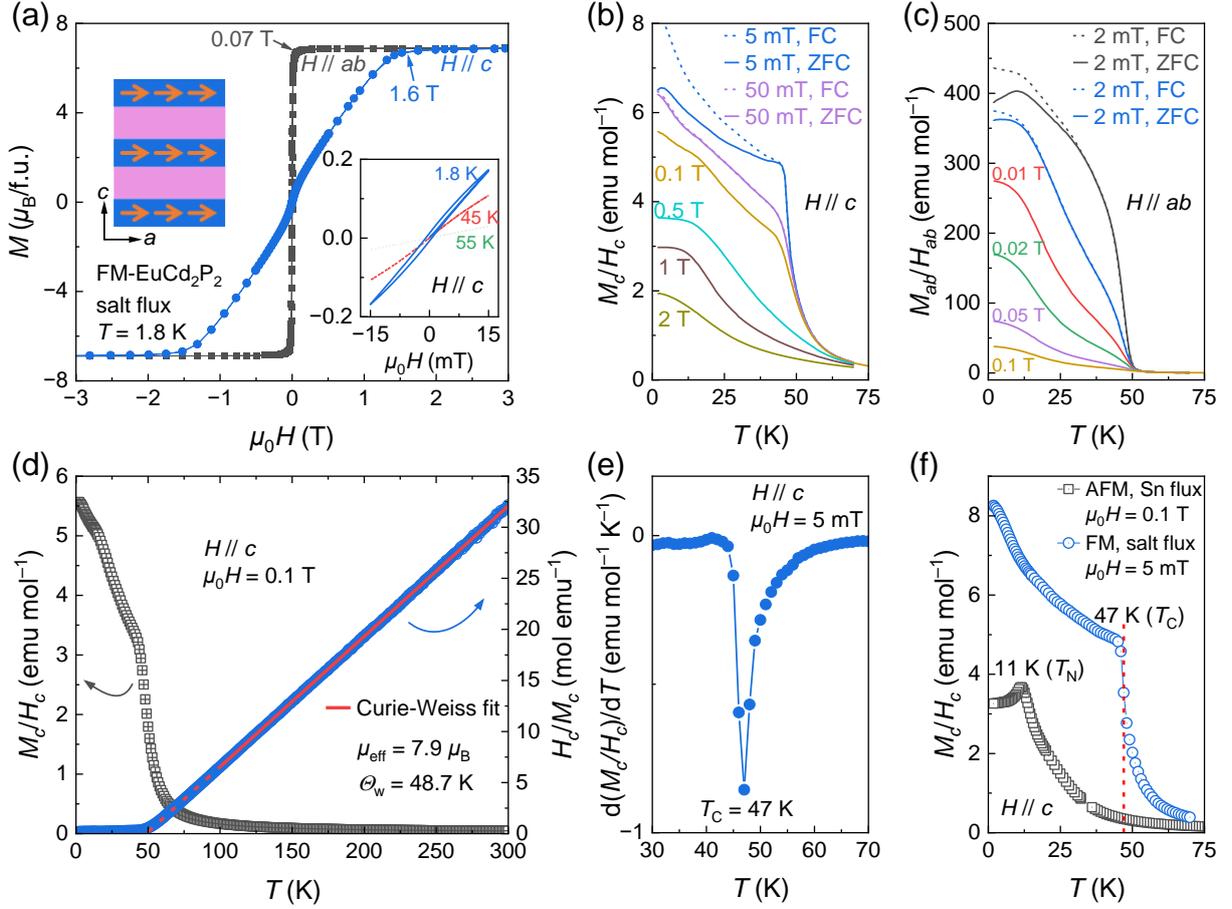

FIG. 2. (a) Magnetization of EuCd$_2$P$_2$ (salt flux) as a function of both in-plane (black squares) and out-of-plane (blue circles) fields at 1.8 K. The upper left schematic illustrated the FM order with in-plane spins. Magnetic hysteresis loops at 1.8 K (blue), 45 K (red), and 55 K (green) are shown in the lower right inset. (b, c) Temperature dependences of magnetic susceptibility ($M/H$) in various out-of-plane (b) and in-plane (c) fields. (d) $M_c/H_c$ measured under 0.1 T ($H // c$) is shown using the left axis, while the right axis corresponds to the reciprocal ($H_c/M_c$) as well as the Curie-Weiss fit from 100 to 300 K. (e) d($M_c/H_c$)/d$T$ as a function of temperature at $\mu_0 H_c$ = 5 mT. (f) Comparison between $M_c/H_c$ of EuCd$_2$P$_2$ with AFM order (black squares, Sn flux) and FM order (blue circles, salt flux).

The susceptibility ($M/H$) further confirm the ferromagnetism of EuCd$_2$P$_2$ (salt flux). The curves with out-of-plane and in-plane fields are plotted in Figs. 2(b) and 2(c). At a small field of 5 mT, $M_c/H_c$ increases noticeably below 60 K and the characteristic bifurcation of zero-field-cooling (ZFC) and field-cooling (FC) data for FM materials is observed below 45 K. The bifurcation is suppressed rapidly with increasing field and becomes indiscernible beyond 0.05 T. The temperature-dependent susceptibility derivative (d($M_c/H_c$)/d$T$) is shown in Fig. 2(e), and the peak temperature of 47 K is taken as the Curie temperature ($T_C$) of FM-EuCd$_2$P$_2$. Figure 2(d) shows the reciprocal of $M_c/H_c$ at 0.1 T (right axis) and the Curie-Weiss fit with $M_c/H_c = \chi_0 + C/(T − \Theta_w)$, which yields $\Theta_w$ = 48.7 K and $C$ = 7.83 emu K mol$^{-1}$. The Weiss temperature $\Theta_w$ agrees excellently with the transition temperature determined by d($M_c/H_c$)/d$T$. The almost identical relation between $\Theta_w$ and $T_C$ indicates that EuCd$_2$P$_2$ is manipulated to be an intrinsic ferromagnet rather than a canted antiferromagnet. From the Curie constant $C$, an effective local moment of $\mu_{eff}$ = 7.9 $\mu_B$ f.u.$^{-1}$ (f.u. denotes formula unit) is obtained, manifesting the Eu$^{2+}$ oxidation state, which is also consistent with the saturated moment of 6.9 $\mu_B$ in Fig. 2(a).

The FM transition is also evident in $M_{ab}/H_{ab}(T)$ curves shown in Fig. 2(c). The susceptibility ascends steeply around 50 K, and the FM bifurcations are observed in small fields as well. The transition temperature determined by d($M_{ab}/H_{ab}$)/d$T$ is identical to the $T_C$ value derived by d($M_c/H_c$)/d$T$, which is shown in Fig. S1 in the SM [22]. Moreover, $\chi_{ab}$ is an order of magnitude larger than $M_c/H_c$ at the same field, which is consistent with the strong magnetocrystalline anisotropy of EuCd$_2$P$_2$. Figure 2(f) compares the $M_c/H_c(T)$ of AFM- and FM-EuCd$_2$P$_2$. The distinct temperature-dependent behaviors unambiguously indicate EuCd$_2$P$_2$ was tuned from an antiferromagnet of $T_N$ = 11 K to a ferromagnet of $T_C$ = 47 K. In addition, the $T_C$ of FM-EuCd$_2$P$_2$ is almost twice the value of FM-EuCd$_2$As$_2$ ($T_C$ = 26 K) [15], as a result of the smaller layer distance (7.177 Å vs. 7.331 Å) and stronger interlayer FM coupling of EuCd$_2$P$_2$.

## C. Electrical resistivity

The temperature and field dependence of charge transport for EuCd$_2$P$_2$ are summarized in Fig. 3. The in-plane resistivity data ($\rho_{xx}$, $I // ab$) collected with out-of-plane field ($H // c$) are presented in Fig. 3(a). Without applying the field, $\rho_{xx}$ shows a typical metallic behavior above 100 K, followed by a modest rise due to magnetic fluctuations. A resistivity peak is located at 47 K, which agrees with $T_C$ determined by the magnetic measurements. Finally, $\rho_{xx}$ decreases monotonically below $T_C$. We notice that the slope of $\rho_{xx}(T)$ below $T_C$ is not uniform, resulting in a hump at around 30 K, which is not observed for FM-EuCd$_2$As$_2$ or other Eu-based layered materials. This feature of $\rho_{xx}$ couldn't be attributed to issues with data collection for reproducibility in different measurements or with different samples. A possible explanation for the hump at 30 K is dimension crossover resulting from enhanced interlayer coupling with decreasing temperature. For AFM-EuCd$_2$P$_2$, strong magnetic fluctuations above $T_N$ (11 K) due to the small spatial extension of P orbitals lead to a colossal resistivity peak at about 1.5$T_N$ (shown in Fig. 3(b), empty squares) [13]. Since the structural parameters are close for FM- and AFM-EuCd$_2$P$_2$, it is plausible to assume competition between short-range magnetic fluctuations and interlayer FM coupling even below the ordering temperature. $\rho_{xx}(T)$ behaviors of AFM- and FM-EuCd$_2$P$_2$ are compared in Fig. 3(b). AFM-EuCd$_2$P$_2$ has about five times the resistivity magnitude of FM-EuCd$_2$P$_2$ at 150 K, and the difference exceeds 300 times at 18 K, the temperature for the maximum CMR effect of AFM-EuCd$_2$P$_2$. Note that this sharp distinction in $\rho(T)$ behavior is caused by the minor Eu vacancies in FM-EuCd$_2$P$_2$.

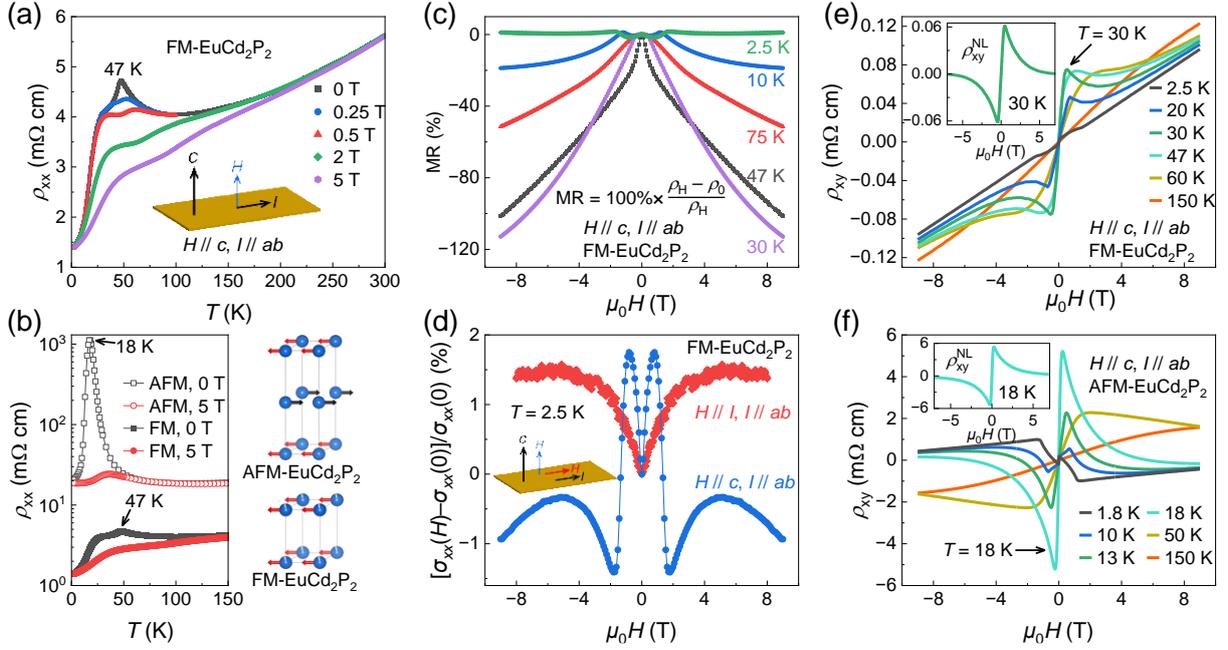

FIG. 3. (a) Temperature dependence of longitudinal resistivity ($\rho_{xx}$) of FM-EuCd$_2$P$_2$ at several fields. (b) Comparison of $\rho_{xx}(T)$ behaviors of AFM- and FM-EuCd$_2$P$_2$, whose speculated magnetic unit cell based on the magnetization are shown on the right. Note that the resistivity is plotted on a logarithmic scale. (c) MR of FM-EuCd$_2$P$_2$ as a function of field at several representative temperatures. (d) Field dependences of magnetoconductivity $[\sigma_{xx}(H) - \sigma_{xx}(0)]/\sigma_{xx}(0)$ at 2.5 K for field parallel to (red) and perpendicular to (blue) the current, respectively. (e) Hall resistivity ($\rho_{xy}$) of FM-EuCd$_2$P$_2$ as a function of field collected between 2.5 K and 150 K. NLAHE at 30 K is extracted and plotted as a function of field in the inset. (f) $\rho_{xy}(H)$ of AFM-EuCd$_2$P$_2$ at different temperatures. Inset shows the NLAHE at 18 K.

After applying the field, both the resistivity peak and hump in Fig. 3(a) are rapidly suppressed. We define the magnetoresistance (MR) as MR = 100% × $[\rho(H) - \rho(0)]/\rho(H)$ and present MR as a function of field at several temperatures in Fig. 3(c). The negative MR starts from high temperature (>150 K) and reaches a maximum of −110% at 30 K under a 9 T field, i.e., a resistivity reduction by half. The negative MR weakens with decreasing temperature below 30 K. At 2.5 K, MR first drops a little in the low-field regime ($\mu_0 H < 0.8$ T), which may be related to the decrease in spin scattering. Then MR increases and becomes almost constant above 1.6 T, indicating that the increased canting of spins towards the $c$ axis in the external field gives rise to the enhancement in resistivity. We also measured the magnetoconductivity at 2.5 K with the field along the current direction ($H // I$, $I // ab$) and compare it to the data with out-of-plane field ($H // c$, $I // ab$) in Fig. 3(d). Like FM-EuCd$_2$As$_2$, positive and negative magnetoconductivity are observed in strong fields for $H // I$ and $H // c$, respectively. In EuCd$_2$As$_2$ and EuCd$_2$Sb$_2$, this difference is usually considered to be related to the chiral anomaly, thus supports the existence of Weyl fermions [16,23]. However, this cause should be excluded on FM-EuCd$_2$P$_2$ for its topologically trivial band structure (shown in Fig. 4(a)). Hence, this result may be ascribed to the magnetism rather than the topological origin.

To further analyze the difference between FM- and AFM-EuCd$_2$P$_2$, we exhibit their Hall resistivity ($\rho_{xy}$) in Figs. 3(e) and 3(f). For FM-EuCd$_2$P$_2$, $\rho_{xy}$ depends on the field almost linearly when the temperature is well above the ordering temperature, e.g., at 150 K. As the temperature decreases, $\rho_{xy}(H)$ becomes nonlinear due to increasing contributions from the anomalous Hall effect. Generally, the total Hall resistivity can be expressed as a sum of three parts, $\rho_{xy} = R_0\mu_0 H + R_S M + \rho_{xy}^{NL}$, where $R_0\mu_0 H$ is linear in the field ($\mu_0 H$) and represents the ordinary Hall effect (OHE), while $R_S M$ is linear in the magnetization ($M$) and represents the conventional anomalous Hall effect (AHE). The third term, $\rho_{xy}^{NL}$, is not linear in either $\mu_0 H$ or $M$ and represents the nonlinear AHE (NLAHE), which is characterized by the peaks in $\rho_{xy}(H)$ curves. $\rho_{xy}^{NL}$ varies strongly with temperature and achieves maximum amplitude around 30 K at 0.5 T, which is shown in the inset of Fig. 3(e) after subtracting the OHE and conventional AHE contributions. The details of extracting NLAHE are provided in Fig. S2 (see SM [22]). At the peak position, the NLAHE constitutes 81% of the total Hall resistivity ($\rho_{xy}^{NL}/\rho_{xy} = 0.81$). Regarding AFM-EuCd$_2$P$_2$, the NLAHE ratio even reaches 97% at 0.25 T for the curve of 18 K. The near-limit ratio indicates that the Hall resistivity of AFM-EuCd$_2$P$_2$ is completely dominated by the NLAHE at the peak position, as in the case of EuCd$_2$As$_2$ [24], which is larger than the values reported in the most NLAHE systems [25–28]. The giant NLAHE in EuCd$_2$P$_2$ may originate from the evolution of the electronic structure induced by the external magnetic field, as observed in EuCd$_2$As$_2$. Future explanatory theories are needed to fully understand the Hall response of EuCd$_2$P$_2$. The AHE contribution has almost vanished above 100 K, and the slopes of OHE contribution don't vary considerably with the temperature. Thus, the carrier concentration of FM-EuCd$_2$P$_2$ is estimated to be ~$4.6 \times 10^{19}$ cm$^{-3}$ (0.0053 hole/f.u. or 0.0026 Eu vacancy/f.u.) at 150 K based on the single-band model, consistent with the negligible Eu vacancies from the structural analysis. By contrast, the hole density of AFM-EuCd$_2$P$_2$ is calculated to be ~$3.6 \times 10^{18}$ cm$^{-3}$ at 150 K (0.0002 Eu vacancy/f.u.), only a tenth of the value of FM-EuCd$_2$P$_2$. The comparison of carrier concentrations manifests that the salt flux method induces a tiny amount of Eu vacancies to EuCd$_2$P$_2$, which leads to the huge changes in the magnetic ground state and transport properties.

**D. DFT calculations**

To gain additional insight into the magnetism of EuCd$_2$P$_2$, we calculated the band structures and total energies for EuCd$_2$P$_2$ with different spin configurations. The band structure for FM-EuCd$_2$P$_2$ (Eu spins along in-plane axis [100]) with spin-orbit coupling (SOC) is displayed in Fig. 4(a), implying that EuCd$_2$P$_2$ is a narrow-gap semiconductor with the direct gap of 380 meV at the Γ point, without considering Eu vacancies. The Eu 4$f$ states can be clearly seen as flat bands located between 1.0 and 1.5 eV below the Fermi level ($E_F$), which also contribute appreciably to the cone-shaped band at the Γ point. Therefore, the cone-shaped hole band is dominated by the strongly hybridized P 3$p$ and Eu 4$f$ orbitals, as well as a smaller contribution from Cd 4$d$ states, as revealed by the density of states (DOS) in Fig. 4(b). The calculations of the band structure with and without SOC indicate that only Eu 4$f$ states are strongly affected, and the corresponding energy gaps are comparable. Besides, the band structure does not vary significantly with the configuration of ground states. The results are presented in Table I and Fig. S3 in the SM [22]. Here we would like to point out that the bad metal for AFM-EuCd$_2$P$_2$ implied by the experiments is inconsistent with the narrow-gap semiconductor implied by the band structures without Eu vacancy in Fig. S3 (see SM [22]), which has also been reported for AFM-EuCd$_2$As$_2$, EuZn$_2$As$_2$, and EuZn$_2$P$_2$ and is likely due to slight defects or disorder in the material [29–33], such as 0.0002 Eu vacancy/f.u. in AFM-EuCd$_2$P$_2$ mentioned earlier.

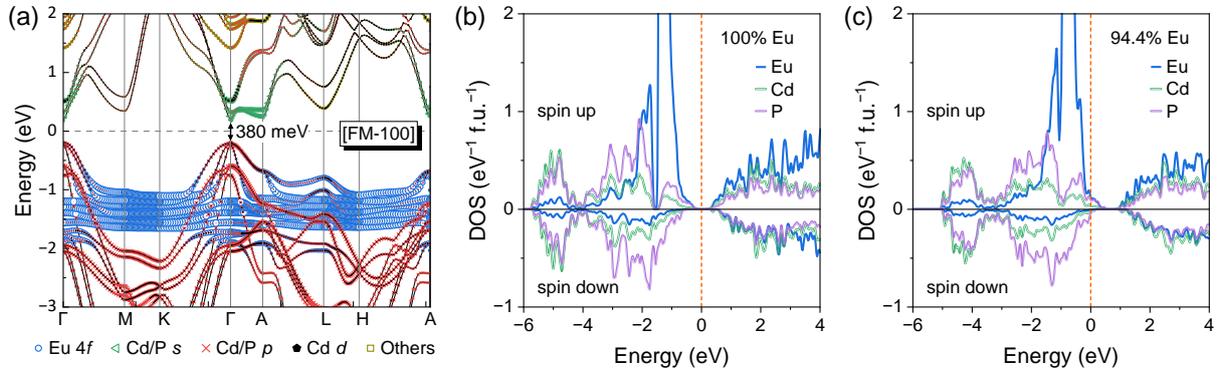

FIG. 4. (a) Orbitally resolved electronic band structure for FM-EuCd$_2$P$_2$ with SOC. (b, c) DOS of in-plane FM-EuCd$_2$P$_2$ calculated with 100% Eu occupation (b) and 94.4% Eu

occupation (c). The Eu vacancy effects was simulated in a 3 × 3 × 2 supercell (Eu concentration ~94.4%), and SOC was not taken into consideration to reduce the computing consumption.

TABLE I. Energy band gaps and energy differences between A-type AFM and FM states obtained by density functional theory (DFT) calculations.

|  | AFM-[100][a] 100% Eu | FM-[100] 100% Eu | AFM w/o SOC 100% Eu | FM w/o soc 100% Eu | AFM w/o soc 94.4% Eu | FM w/o soc 94.4% Eu |
| --- | --- | --- | --- | --- | --- | --- |
| Energy gap [meV] | 550 | 380 | 590 | 410 | / | / |
| $E_{FM} - E_{AFM}$ [meV/f.u.] | / |  | 0.29 |  | −2.02 |  |

[a]A-type antiferromagnetic, spins aligned along [100].

Next, we examine the correlation between the Eu vacancies and the magnetic ground states. For the structure without Eu vacancies, the total energy of the FM order is found to be 0.29 meV higher than the A-type AFM order, in agreement with the AFM order given by the experiments. The small energy difference also suggests the possibility of manipulating the magnetic order of EuCd$_2$P$_2$. We then simulated the effects of Eu vacancy by removing one Eu atom in a 3 × 3 × 2 supercell, resulting in 5.6% vacancies. The energy of the FM state is 2.02 meV lower than the A-type AFM order, indicating that the interlayer FM coupling is more favored in the presence of Eu vacancies. In a recent study, the interlayer AFM coupling of the sister material EuZn$_2$P$_2$ was explained as realized by the superexchange interaction through the Eu-P-P-Eu path, with an exchange parameter of –0.35 meV [34]. Since this conclusion is likely to be appliable to EuCd$_2$P$_2$, it is plausible to assume that the interlayer AFM interaction remains the same after considering the vacancy effects for EuCd$_2$P$_2$, due to the minor structural modification. Accordingly, the sign reversal of energy difference demonstrates that the enhanced interlayer FM coupling induced by the carriers surpasses the superexchange interaction. In fact, the ferromagnetism induced by a low carrier concentration is not unusual. Established cases include the FM semiconductors Eu$X$ ($X$ = O, S, Se, Te) [9,35], diluted magnetic semiconductor (Ga,Mn)As [8,36], Mn pyrochlore Tl$_{2-x}$Sc$_x$Mn$_2$O$_7$ [37], and other low-carrier-density ferromagnets like Ca$_{1-x}$La$_x$B$_6$, EuB$_6$, UTeS, PbSnMnTe, etc [38–41]. The FM states with transitions below 20 K are also reported by slight chemical doping in EuCd$_2$As$_2$ [42]. Since the indirect exchange mediated by a low carrier concentration (<10$^{21}$ cm$^{-3}$) cannot be understood within the framework of the Ruderman-Kittel-Kasuya-Yosida (RKKY) interaction [43], further efforts will be devoted to investigating the FM behavior and the dependence of $T_C$ in FM-EuCd$_2$P$_2$.

### E. Magnetic state manipulation

We have demonstrated that the magnetic ground state of $EuCd_2P_2$ can be tuned from an AFM state of $T_N = 11$ K into a FM state of $T_C = 47$ K by a low hole density ($<10^{20}$ cm$^{-3}$), accompanied by the giant differences in the transport properties. However, the carrier density induced by defects or chemical doping cannot be continuously tuned. The gating technique based on field-effect transistors (FETs) enables the precise modulation of carrier density in the materials, making it applicable to thin films or micrometer-scale single-crystalline lamellae of $EuCd_2P_2$ [44−47]. Figures 5(a) and 5(b) show schematics of the primary AFM-$EuCd_2P_2$ and FM-$EuCd_2P_2$ born of Eu defects, while Fig. 5(c) illustrates the idea of tuning the magnetic state of $EuCd_2P_2$ with the gating method. Through electric-field-induced electrostatic doping, a volume carrier density as high as $10^{21}$ cm$^{-3}$ can be achieved [48], which is enough to induce FM order in $EuCd_2As_2$ and $EuCd_2P_2$. If reversible magnetism between AFM-$EuCd_2P_2$ and FM-$EuCd_2P_2$ is realized by controlling the electric field, spintronics utilizing the CMR effect of AFM-$EuCd_2P_2$ could be promising. Moreover, not only is a higher $T_C$ probable with optimized carrier density [49], but also a "real" semiconducting state of AFM-$EuCd_2P_2$ could be obtained by reducing the carrier density [29]. Additionally, continuous electric tuning of magnetization dynamics may lead to novel quantum phenomena at the critical points, such as the topological features.

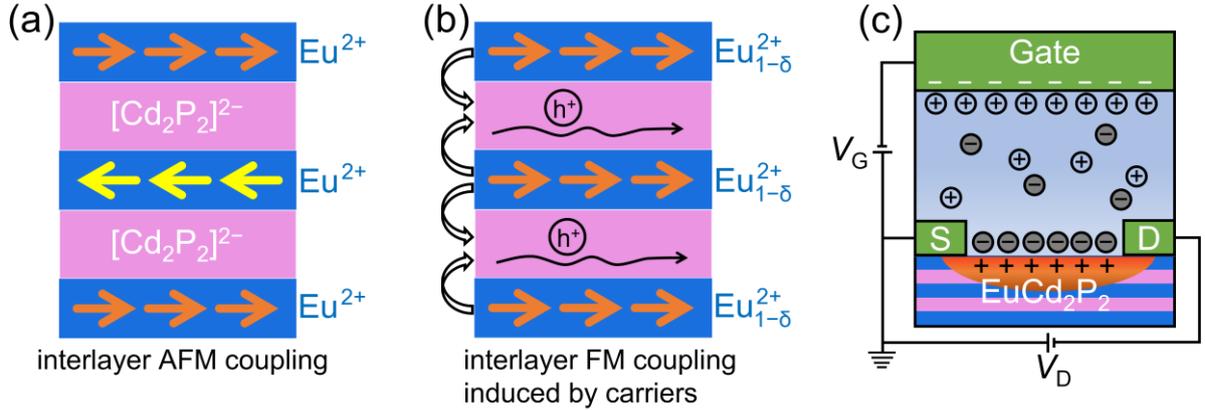

FIG. 5. (a) Simplified schematic of the magnetic structure of AFM-$EuCd_2P_2$. (b) Illustration of interlayer FM coupling induced by hole doping resulting from Eu vacancies. (c) Idea of tuning the magnetic state of $EuCd_2P_2$ with an electric-double-layer transistor (EDLT). $V_G$ and $V_D$ denote gate voltage and drain voltage, respectively.

### IV. CONCLUSION

In summary, we successfully synthesized single crystals of FM-EuCd$_2$P$_2$ with an ordering temperature of 47 K. FM-EuCd$_2$P$_2$ exhibits metallic behavior with a modest negative MR effect in the field, in sharp contrast to AFM-EuCd$_2$P$_2$ with a CMR effect. The significant change in the magnetic ground state and transport properties could be attributed to the Eu vacancies resulting from the molten salt flux method. DFT calculations shows that the small number of Eu defects enhances interlayer FM coupling, and thus the FM order is favored. As the properties of EuCd$_2$P$_2$ could be altered by a low carrier concentration, we propose manipulating the state of EuCd$_2$P$_2$ with the gating technique, which can control the carrier concentration quickly, continuously, and reversibly. The strongly coupled magneto-electronic properties, as well as the high tunability, make EuCd$_2$P$_2$ a potential candidate for future electric-field-controlled spintronic devices.

In addition, it is of great interest to explore other methods for tuning the properties of EuCd$_2$P$_2$, such as pressure, strain, chemical doping, and light [42, 50−53], which are effective ways to induce carriers or change electronic band topology. Moreover, the diverse Eu-based layered Zintl compounds, including EuIn$_2$As$_2$, EuZn$_2$As$_2$, EuMn$_2$P$_2$, and others [54,55], with an AFM ground state and a narrow energy gap, provide a golden opportunity to tune the magnetism and find exotic physical properties by controlling the carrier density in the materials.

**ACKNOWLEDGMENTS**

This work was supported by the National Natural Science Foundation of China (Grants No. 12204094), the Natural Science Foundation of Jiangsu Province (Grant No. BK20220796), the Start-up Research Fund of Southeast University (Grant No. RF1028623289), the Interdisciplinary program of Wuhan National High Magnetic Field Center (WHMFC) at Huazhong University of Science and Technology (Grant No. WHMFC202205), and the open research fund of Key Laboratory of Quantum Materials and Devices (Southeast University), Ministry of Education.

**Supplemental Materials: Manipulating magnetism and transport properties of EuCd$_2$P$_2$ with a low carrier concentration**

Xiyu Chen,[1] Ziwen Wang,[1] Zhiyu Zhou,[1] Wuzhang Yang,[2,3] Yi Liu,[4] Jia-Yi Lu,[5] Zhi Ren,[2,3] Guang-Han Cao,[5,6] Fazel Tafti,[7] Shuai Dong,[1,§] and Zhi-Cheng Wang[1,**]

[1]*Key Laboratory of Quantum Materials and Devices of Ministry of Education, School of Physics, Southeast University, Nanjing 211189, China*

[2]*School of Science, Westlake University, Hangzhou 310024, China*

[3]*Institute of Natural Sciences, Westlake Institute for Advanced Study, Hangzhou 310024, China*

[4]*Department of Applied Physics, Zhejiang University of Technology, Hangzhou 310023, China*

[5]*School of Physics, Interdisciplinary Center for Quantum Information and State Key Laboratory of Silicon and Advanced Semiconductor Materials, Zhejiang University, Hangzhou 310058, China*

[6]*Collaborative Innovation Centre of Advanced Microstructures, Nanjing University, Nanjing 210093, China*

[7]*Department of Physics, Boston College, Chestnut Hill, Massachusetts 02467, USA*

**A. Structural analysis for FM-EuCd$_2$P$_2$**

Table S1 provides the refinement results for the single-crystal diffraction data of FM-EuCd$_2$P$_2$ collected with the Mo $K\alpha$ radiation at 150 K. The occupancies of Cd and P were fixed to be 1.0 to avoid unphysical values greater than 1. The unit cells and atomic coordinates of FM-EuCd$_2$P$_2$ (150 K) and AFM-EuCd$_2$P$_2$ (298 K) are listed in Table S2 for comparison. The atomic coordinates of Cd ($z_{Cd}$) and As ($z_{As}$) are almost the same within experimental error for both materials. And the shrinkage of unit cells for FM-EuCd$_2$P$_2$ is due to the low temperature for data collection, manifested by the close ratios of $c/a$. CCDC 2313633 contains a full set of crystallographic data for FM-EuCd$_2$P$_2$.

§ sdong@seu.edu.cn
** wzc@seu.edu.cn

Table S1. Crystallographic data and refinement results for a single crystal of FM-EuCd$_2$P$_2$ (salt flux) at 150 K.

|  | FM-EuCd$_2$P$_2$ |
| --- | --- |
| Space group | $P\bar{3}m1$ (No. 164) |
| $a$ (Å) | 4.3168(2) |
| $c$ (Å) | 7.1616(5) |
| $V$ (Å$^3$) | 115.575(13) |
| $Z$ | 1 |
| $T$ (K) | 150 (2) |
| Eu site occupancy | 0.998(5) |
| Cd site occupancy | 1 |
| P site occupancy | 1 |
| $U_{Eu}$ (Å$^2$) | 0.0199(8) |
| $U_{Cd}$ (Å$^2$) | 0.0197(8) |
| $U_{P}$ (Å$^2$) | 0.0191(12) |
| GoF | 1.368 |
| $R_1(F)$[a] | 0.0376 |
| $wR_2(F^2)$[b] | 0.1247 |

[a] $R_1 = \Sigma ||F_o| - |F_c||/\Sigma |F_o|$
[b] $wR_2 = [\Sigma w(F_o^2 - F_c^2)^2/\Sigma w(F_o^2)^2]^{1/2}$

Table S2. Comparison between the unit cells and atomic coordinates of FM-EuCd$_2$P$_2$ (150 K) and AFM-EuCd$_2$P$_2$ (298 K). The space group is $P\bar{3}m1$ (No. 164), and the Wyckoff sites are Eu 1$a$ (0,0,0), Cd 2$d$ (1/3,2/3,$z$), and P 2$d$ (1/3,2/3,$z$). Note that no Eu vacancies were identified in AFM-EuCd$_2$P$_2$ samples.

|  | FM-EuCd$_2$P$_2$ (150 K) | AFM-EuCd$_2$P$_2$ (298 K) |
| --- | --- | --- |
| $a$ (Å) | 4.3168(2) | 4.3248(2) |
| $c$ (Å) | 7.1616(5) | 7.1771(7) |
| $c/a$ | 1.6590 | 1.6595 |
| $z_{Cd}$ | 0.6356(3) | 0.6357(1) |
| $z_{As}$ | 0.2486(6) | 0.2484(1) |

## B. Susceptibility and Curie-Weiss analysis

Susceptibility ($M/H$) of FM-EuCd$_2$P$_2$ measured under a small field of 2 mT in the *ab* plane is shown in Fig. S1(a). Fig. S1(b) shows the transition temperature (47 K) determined by the derivative d($M_{ab}/H_{ab}$)/d$T$, which is identical to the $T_C$ value derived by d($M_c/H_c$)/d$T$ in the main text. Figure S1(c) shows the Curie-Weiss (CW) analysis with the in-plane field of 0.1 T. The right *y* axis corresponds to $1/(M_{ab}/H_{ab} - \chi_0)$ (blue circles), where $\chi_0$ is a small background. The red solid line show the CW fit according to $M_{ab}/H_{ab} - \chi_0 = C/(T - \Theta_W)$ from 100 to 300 K, and the red dash line is the linear extension. The Weiss temperature $\Theta_W$ and the effective moment $\mu_{eff}$ are consistent with the fitting results from the $M_c/H_c$ data (Fig. 2(d) in main text).

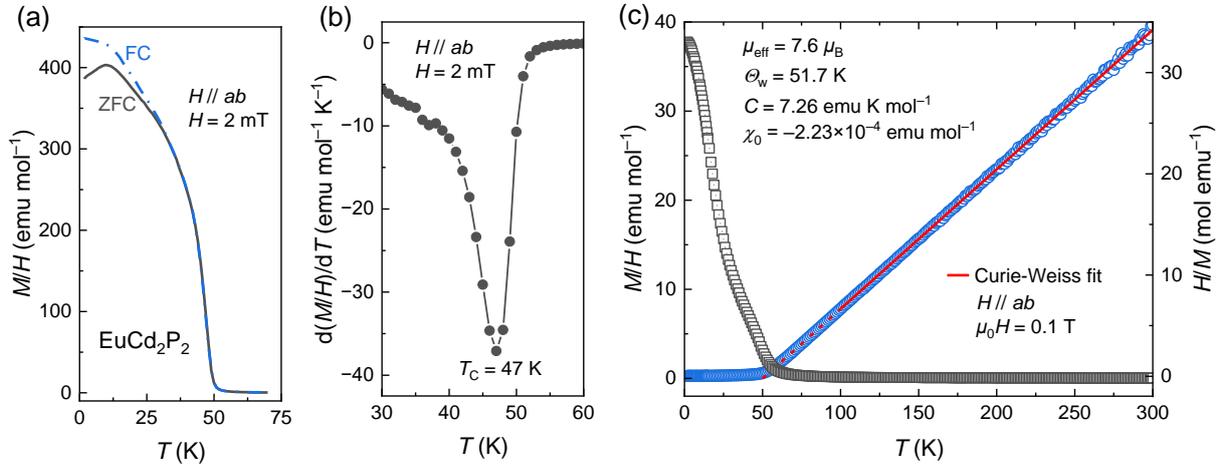

FIG. S1. (a) Zero-field-cooling (black) and field-cooling (blue) data under the in-plane field of 2 mT for FM-EuCd$_2$P$_2$. (b) d($M_{ab}/H_{ab}$)/d$T$ as a function of temperature at $H_{ab}$ = 2 mT. (c) The left axis is the susceptibility measured under 0.1 T ($H // ab$), while the right axis corresponds to the reciprocal of susceptibility as well as the Curie-Weiss fit from 100 to 300 K.

## C. Decomposition of the field-dependent Hall resistivity

We extract the nonlinear anomalous Hall effect (NLAHE) by removing ordinary Hall effect (OHE) and conventional anomalous Hall effect (AHE) from the total Hall resistivity. The total Hall resistivity is expressed as $\rho_{xy} = R_0\mu_0H + R_SM + \rho_{xy}^{NL}$. The OHE contribution is linear in the field $\mu_0H$ and the conventional AHE contribution is proportional to the magnetization $M$. The magnetization varies mildly when the field is high, so that can be taken as a constant value. Then $R_0\mu_0H + R_SM$ is only dependent on $\mu_0H$ for high fields. We fit the original Hall resistivity $\rho_{xy}(H)$ from 6 T to 7 T to the equation $R_0\mu_0H + R_SM$, where $R_0$ is the resulting slope, and $R_S$ could be obtained from the intercept $R_SM$. At 30 K, the resultant $R_0$ and $R_S$ for FM-EuCd$_2$P$_2$ are 8.351 μΩ cm/T and 6.098 μΩ cm/$\mu_B$. While at 18 K, $R_0$ and $R_S$ for AFM-EuCd$_2$P$_2$ are 68.26 μΩ cm/T and −137 μΩ cm/$\mu_B$.

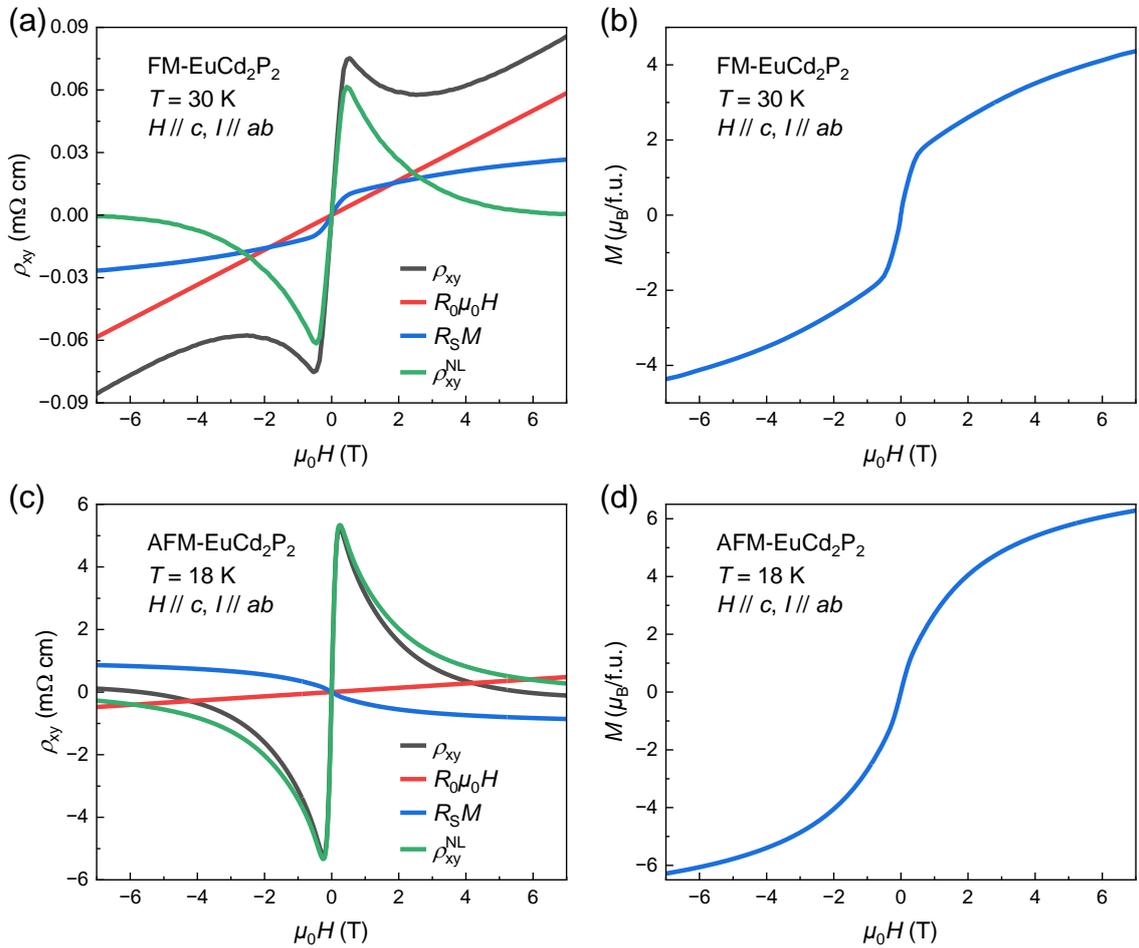

FIG. S2. (a) Decomposition of Hall resistivity of FM-EuCd$_2$P$_2$ at 30 K. The total Hall resistivity, OHE, conventional AHE, and NLAHE are plotted as black, red, blue, and green curves, respectively. (b) Magnetization as a function of field ($H // c$) at 30 K for FM-EuCd$_2$P$_2$. (c) Decomposition of Hall resistivity of AFM-EuCd$_2$P$_2$ at 18 K. (d) $M(H)$ curve ($H // c$) at 18 K for AFM-EuCd$_2$P$_2$.

## D. Band structure calculations

The band structures calculated with different configurations are shown in Fig. S3, with the energy gaps between 0.38 and 0.59 eV. The band structure does not vary significantly from ferromagnetic (FM) order to antiferromagnetic (AFM) order. When the spin-orbit coupling (SOC) considered, the flat bands of Eu 4$f$ states is decoupled around binding energy 1.3 eV.

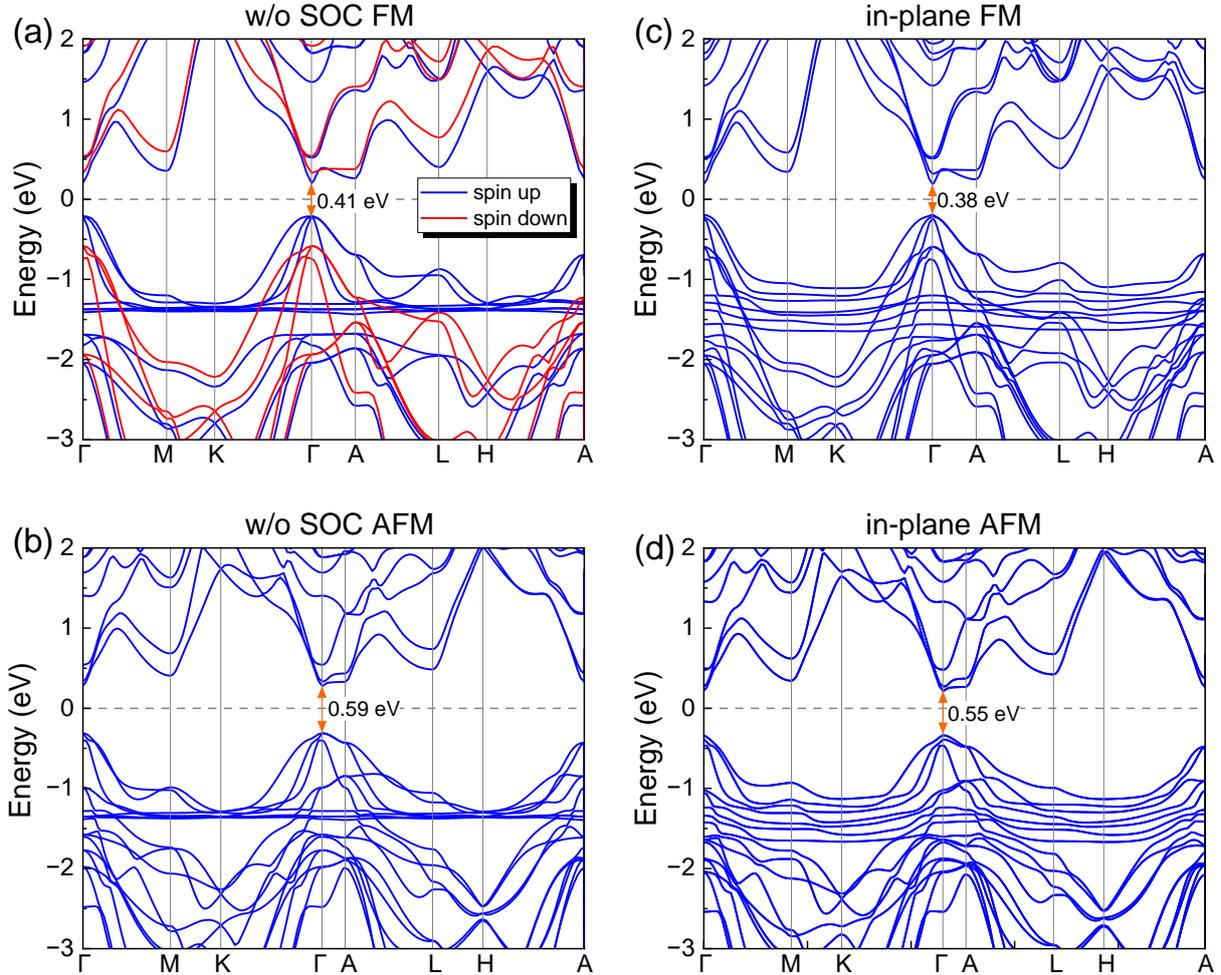

FIG. S3. The band structures calculated with different configurations. (a) FM, without SOC; (b) AFM, without SOC; (c) in-plane FM; (d) in-plane AFM.

Since the in-plane magnetic order may break the threefold symmetry, we calculated and compared the band structures with the magnetic order along the in-plane axis [100] and [010], respectively. As shown in Fig. S4, the energy bands with the two spin moment axes are almost identical at the valence band maximum and the conduction band minimum, while show finite differences in the relatively deep energy level (−1.5 ~ −1 eV).

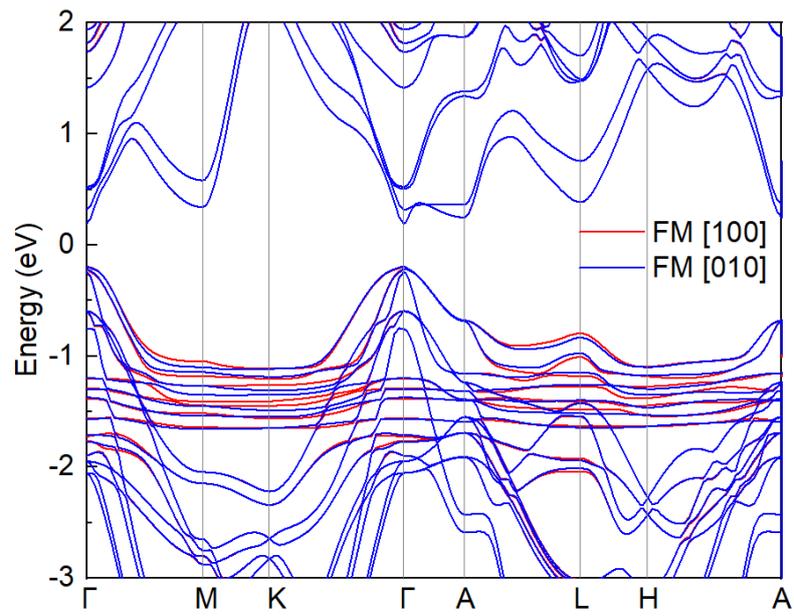

Fig. S4 Band structures with the magnetic order along the in-plane axis [100] and [010].